\documentclass[aps,pre,twocolumn,groupedaddress,floatfix,showpacs]{revtex4}

\usepackage{graphicx}
\usepackage{latexsym}

\begin{document}

\begin{widetext}
\noindent\textbf{Preprint of:}\\
Wolfgang Singer, Timo A. Nieminen,
Norman R. Heckenberg and Halina Rubinsztein-Dunlop\\
``Collecting single molecules with conventional optical tweezers''\\
\textit{Physical Review E} \textbf{75}(1), 011916 (2007)

\vspace{2mm}

~
\end{widetext}

\title{Collecting single molecules with conventional optical tweezers} 

\author{Wolfgang Singer}
\email[]{singer@physics.uq.edu.au}
\author{Timo A. Nieminen}
\email[]{timo@physics.uq.edu.au}
\author{Norman R. Heckenberg}
\author{Halina Rubinsztein-Dunlop}

\affiliation{Centre for Biophotonics and Laser Science, Department of Physics,
The University of Queensland, Brisbane QLD 4072, Australia}

\begin{abstract}
The size of particles which can be trapped in optical tweezers ranges
from tens of nanometres to tens of micrometres. This size regime also
includes large single molecules. Here we present experiments
demonstrating that optical tweezers can be used to collect polyethylene
oxide (PEO) molecules suspended in water. The molecules that accumulate
in the focal volume do not aggregate and therefore represent a region of
increased molecule concentration, which can be controlled by the
trapping potential. We also present a model which relates the change in
concentration to the trapping potential. Since many protein molecules
have molecular weights for which this method is applicable the effect
may be useful in assisting nucleation of protein crystals.
\end{abstract}

\maketitle 

Optical tweezers can trap and manipulate micrometer sized particles, and
a variety of applications in physics, chemistry and biology have been
explored~\cite{ashkin1970,grier2003}. The underlying principle is
photon momentum transfer which, for a tightly focussed laser beam,
results in the creation of a three-dimensional trapping
potential~\cite{ashkin1986}. The depth of the optically induced
trapping potential is determined by, among other parameters, the
polarizability of the trapped object and the intensity of the laser
beam. The polarizability of a Rayleigh particle scales with its
volume~\cite{svoboda1994}. Since trapping of a particle requires
a trapping potential which is sufficiently larger than the thermal
energy of the particle, there is a minimum size of particle that can
be trapped. Trapping of solid particles with sizes down to tens of
nanometres has been reported~\cite{svoboda1994,harada1996}. Also
aggregation of polymer chains with radii of gyration in this size
regime in an optical field has been observed~\cite{smith1999}. However,
there are important thermodynamic differences between the trapping of
particles that aggregate and particles that do not---in the latter
case, the partial pressure due to the trapped particles increases as
the concentration increases, while in the former, the number density does
not increase. The trapping of large macromolecules has been
demonstrated~\cite{katsura1998,ichikawa2005}, in this case DNA molecules of
$\approx 100,000$\,kDa, but there do not seem to have been
any studies of the trapping of non-aggregrating molecules of sizes
spanning the limits of what can be trapped, which is orders of magnitude
smaller than this.

Here we report on experiments carried out with PEO molecules of
different molecular weight (Sigma Aldrich, USA, $m_w=100\,$kDa,
300\,kDa, 900\,kDa). We show that the concentration of molecules
within the trapping region can be controlled and reversibly changed
by the trapping power. We also show that only molecules above a certain
molecular weight can be trapped.
 
\begin{figure}[!htbp]
\centerline{\includegraphics[width=0.97\columnwidth]{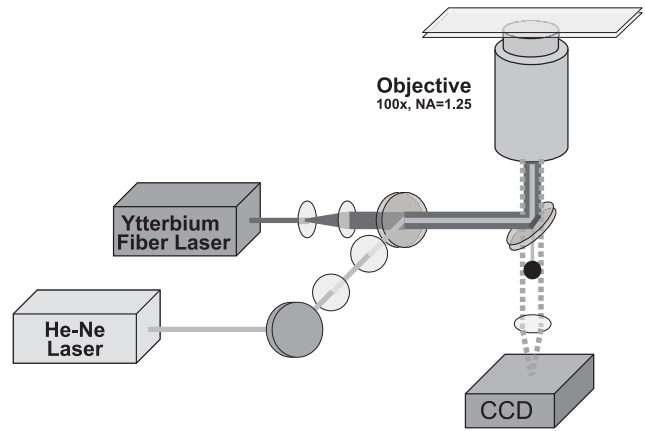}}
\caption{Schematic diagram of the experimental setup. The ytterbium fiber laser was used to create the trapping potential while the local increase in the molecule concentration was monitored by collecting the scattered light of the collinear He-Ne laser. A beam block in the detection beam path increased the sensitivity of the detection system.}
\label{setup}
\end{figure}

To create the trapping potential we used a standard tweezers setup,
based on an inverted microscope (see figure 1), where the CW trapping
laser (ytterbium fiber laser, 1064\,nm, IPG, USA) was coupled into a
$100\times$, NA=1.25 objective (Olympus, Japan) and brought to a focus
in the liquid containing the PEO molecules. The power of the trapping
laser was varied between 0 and 0.7\,W in the object plane. The temperature
of the sample slide and the objective were kept constant using
thermoelectric coolers.

To monitor the effects of the trapping potential on the PEO molecules a low power He-Ne laser (JDS Uniphase, USA) was aligned collinearly with the trapping laser beam, and focused onto the same position. The increased concentration of the initially homogeneously dispersed molecules in the focal region is associated with a higher index of refraction -- as compared to the surrounding water -- and thus resulted in scattering of the He-Ne light. To further increase the sensitivity of the setup we blocked the central portion of the He-Ne reflection, allowing only rays which are scattered at angles larger than those of the incident beam to reach the CCD camera. For this purpose the incident He-Ne laser beam did not fill the back aperture of the objective, thus a beam block between the dichroic mirror and the CCD camera could be used to discriminate between light reflected from the glass surfaces and the light scattered from the probe volume. The scattered light was monitored with a CCD camera, and its intensity was quantified using a MATLAB program evaluating the corresponding pixels. 

The PEO was dissolved in deionized water and the whole sample was heated up to $50^\circ$C for several days prior to use to ensure that the molecules were completely dissolved. The radii of gyration of the respective PEO molecules in water together with the respective overlap concentrations and the concentrations actually used are listed in table 1~\cite{zebrowski2003,dasgupta2002,Devanand91}. The different sample solutions all had the same number density of dissolved PEO molecules ($\approx 300 \mu$m$^{-3}$). The corresponding PEO concentrations in the different samples were all well below the overlap concentration, thus the PEO molecules could be treated as single particles. 

\begin{table}
	\centering
		\begin{tabular}{c|c|c|c}
			 molecular  & gyration & overlap conc. & conc. of sample\\  
			 weight [kDa] &  radius [nm] & [wt\,\%] / $\mu$m$^{-3}$ &  [wt\,\%] / $\mu$m$^{-3}$ \\\hline
			 100 & 17.6 \cite{Devanand91} & 0.48 \cite{dasgupta2002} / 15033 & 0.01 / 313 \\ 
			 300 & 33.5 \cite{Devanand91} & 0.36 \cite{zebrowski2003} / 3758 & 0.027 / 281 \\ 
			 900 & 63.6 \cite{Devanand91} & 0.16 \cite{dasgupta2002} / 556 & 0.077 / 268 \\  
		\end{tabular}
	\caption{Radii of gyration, overlap concentrations, and concentrations of the actually used samples for different molecular weight PEO molecules.}
	\label{tab:RadiiOfGyrationOfPEOWithDifferentMolecularWeights}
\end{table}

Since the power of the detection laser was kept constant for all measurements ($P_\mathrm{HeNe}=2$\,mW), the scattered light intensity was a measure of the increase of the concentration in the focal region. A trapping potential created by the fiber laser typically caused an increase of the intensity of the scattered light until it eventually reached a stable value (see figure 2). This value represented the steady-state equilibrium concentration for the given potential. The intensity of the scattered light increased exponentially, as the fit with a function of the type $S=S_0 \cdot (1-e^{-t/\tau})$ showed.

\begin{figure}[!bhtp]
\centerline{\includegraphics[width=0.97\columnwidth]{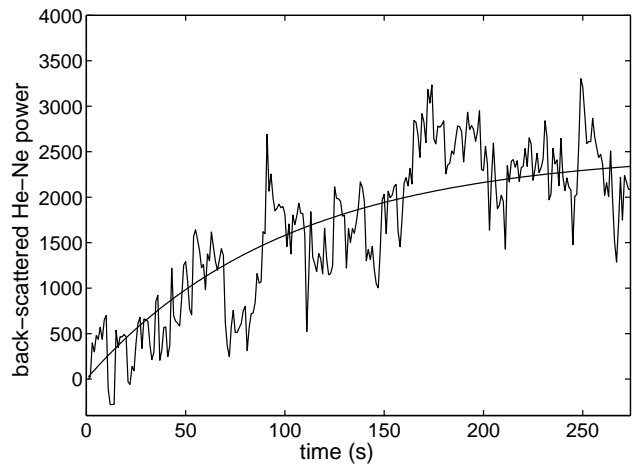}}
\caption{Typical time series of the scattered He-Ne light intensity
(in arbitrary units) as a
result of the concentration increase due to a trapping potential created
by the trapping laser (switched on at t=0). The solid line shows a fit
curve using the  function $S=S_0 \cdot (1-e^{-t/\tau})$.}
\label{exp_increase}
\end{figure}

The typical timescales to reach the steady-state concentration were in the order of minutes, depending on trapping power. The long timescales rule out the possibility that the light scattering is caused by a temperature induced change of the refractive index of the medium, since temperature effects due to heating would occur in milliseconds~\cite{niemz2004}.

To demonstrate that the molecule concentration could be increased reversibly the trapping power was changed stepwise (figure 3(a)). Laser powers above a threshold power resulted in an increase of the concentration of the molecules in the focal volume, whereas trapping laser powers below the respective threshold power led the scattered light intensity to decrease to zero. This threshold behavior is similar to the switch--on behavior of the current--voltage characteristic of a diode. As for diodes, the relation is in fact exponential, which means that for trapping potentials above a certain threshold --- the thermal energy --- the concentration increases dramatically.

The intensity of the scattered He-Ne laser light depends on unknown parameters, like the elastic scattering properties of the PEO molecules. Thus the scattered light can not directly be used to quantify the concentration in the focal volume. Nevertheless, the intensity of the scattered light is a measure of the relative concentration change, and by fitting the experimentally obtained intensities with the mentioned exponential function, time-constants as a function of trapping laser powers could be obtained. The reciprocal of the time-constant $\tau$ can be called a collection rate, which we denote by $R$.

\begin{figure}[!htbp]
\centerline{\includegraphics[width=1.01\columnwidth]{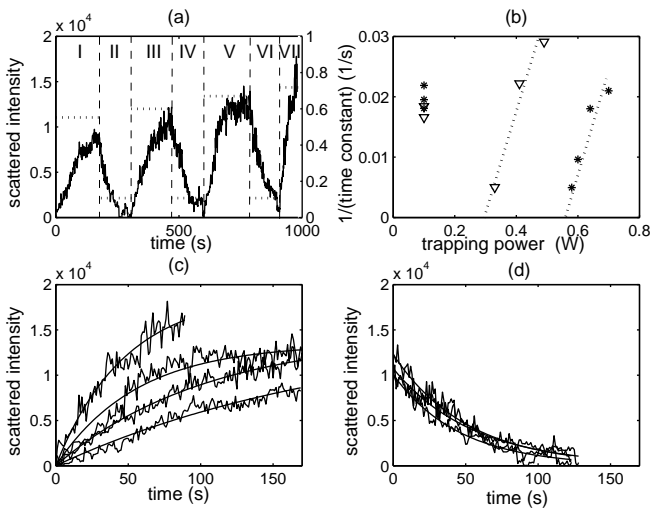}}
\caption{(a) Scattered He-Ne light intensity
(in arbitrary units) for a stepwise changing
power of the trapping laser (dotted lines). Graph is for PEO with
molecular weight $m_w=300\,$kDa and an initial PEO concentration of
0.027 wt\,\%, the trapping laser powers in the different sections
(I-VII) were: 0.58\,W-0.10\,W-0.60\,W-0.10\,W-0.64\,W-0.10\,W-0.7\,W.
(b) Summary of collection rates $R$ for PEO with molecular weights
$m_w=300\,$kDa ($\ast$) and $m_w=900\,$kDa ($\nabla$). The data points
at 0.1W give the reciprocal of the time--constants for the
diffusion--driven decrease extracted from (d). (c) Increase of He-Ne
light intensity for different laser powers (taken from (a)), fitted with
$S=S_0 \cdot (1-e^{-t/\tau})$. (d) For laser powers below the threshold
the He-Ne light intensities decreased exponentially with time. }
\end{figure}


A summary of the collection rates $R$ as a function of the trapping power used for PEO molecules with molecular weight $m_w=300\,$kDa and $m_w=900\,$kDa is plotted in figure 3(b). For laser powers which cause an increase of the concentration the collection rate $R$ increases with the power of the trapping laser (figure 3(c)). From figure 3(b) the threshold trapping power for the respective molecular weight (in this case for $m_w=300\,$kDa) could be extrapolated. We defined the threshold power for the different molecular weight PEO molecules at the intersection of the fitted straight line with the $x$-axis. The threshold for the molecules with $m_w = 900\,$kDa is $P\approx0.29\,$W, and for $m_w = 300\,$kDa is $P\approx0.53\,$W.

As expected, the higher the molecular weight of the PEO molecules the lower the threshold trapping powers needed to confine the molecules. For PEO molecules with a molecular weight of $m_w=100\,$kDa the concentration of the molecules could not be increased to a detectable level (using laser powers of up to 0.7\,W). 

If the laser power was below the threshold for the respective PEO molecules, the decrease of the scattered He-Ne light intensities had the same time-constants (independent of the scattered light intensity at the steady-state concentration). The time-constants for the $m_w=900\,$kDa PEO molecules were slightly larger than the ones for the $m_w=300\,$kDa molecules. This is consistent with theory which predicts smaller diffusion-coefficients for larger particles. This can be attributed to the fact that for laser powers below the threshold the decrease of the concentration of molecules in the focal volume is solely driven by diffusion. Thus the time-constants for the decrease are independent of the initial concentration. 

This is in contrast to the increase of molecule concentration which dependents on the trapping potential, and consequently the time-constant depends on the trapping power.

Our finding that the time-constant depends on the trapping power is consistent with a calculation of the achievable trapping potential for a particle in a laser beam with a Gaussian intensity profile. Since the individual molecules are much smaller than the wavelength of the trapping laser the gradient force 
\begin{equation}
 F_\mathrm{grad} = \frac{\alpha}{2 c_0 n_m \epsilon_0}\nabla \vec{S},   
\end{equation}
can be calculated using Rayleigh scattering, where $\vec{S}$ is the Poynting vector, and the other symbols have their usual meaning~\cite{harada1996,stratton1941}. The quasi-static polarizability $\alpha$ of a non-magnetic dielectric sphere of radius $r$ and refractive index $n$, immersed in a medium of refractive index $n_m$ is given by~\cite{stratton1941} 
\begin{equation}
 \alpha = 4\pi\epsilon_0 n^2_m r^3 \frac{n^2_r -1}{n^2_r +2},   
\end{equation}
where $n_r=n/n_m$ denotes the relative refractive index. A comparison of the gyration radii of PEO in water with the stretched chain lengths \cite{cooper1991} ($l_c\approx2.4\,$$\mu$m for 300\,kDa, and $\approx$ 7.2\,$\mu$m for 900\,kDa) of the PEO molecules shows that the individual molecules can be approximated as spheres with gyration radii $r_g$. A subsequent integration of the gradient force over the beam profile 
\begin{equation}
U(r_0) = - \int^{r_0}_{-\infty} F_\mathrm{grad} dr
\end{equation}
gives the trapping potential. Since the gradient force is proportional to the intensity gradient, the trapping potential at any point is simply proportional to the intensity at this point.

In order to trap a particle the trapping potential needs to be larger than the mean thermal energy $kT$ of the particle. On this account we calculated a contour for which the particle has to overcome a potential of $kT$ to escape the trap. This contour is associated with a certain surface area, and this surface area increases with the trapping power. The increase is due to the fact that for higher laser powers a potential depth of $kT$ is already created at larger distances from the focal point. 

The extension of the `$kT$' surface area with the laser power is reflected by the decreased time-constant necessary to reach the steady-state intensity. This can be illustrated by visualizing the equi-intensity contour as an event horizon, which has a larger surface area for higher laser powers, and a smaller for lower powers. The event horizon divides the liquid into a region where particles can freely diffuse, and a region (the volume inside of the contour) where the particles remain trapped, once they have entered this region (by diffusion). Since the potential of the trap outside of this event horizon is too small to attract particles the trap can only be loaded by passive diffusion of the particles suspended in the surrounding liquid. Particles which diffuse through the mentioned surface area get trapped in the focal volume. Since those particles can no longer diffuse out of the trap, the concentration in the focal volume increases with time. Since the area of the surface determines how many particles per time diffuse through it, it directly determines the initial collection rate. 

If the trapping power is below the threshold, the particles can diffuse out of the trap. The time-constant for the diffusion out of the trap is (for $U_\mathrm{trap}\ll kT$) is only determined by the diffusion coefficient of the particles in the liquid. 


\begin{figure}[!htbp]
\centerline{\includegraphics[width=0.97\columnwidth]{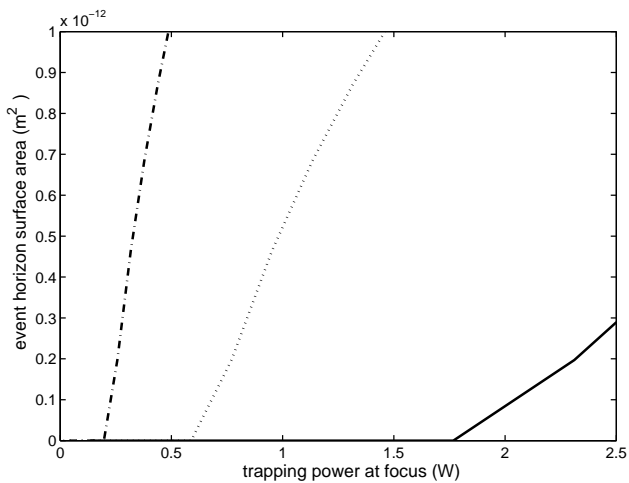}}
\caption{Calculated surface area of the equi-potential contour of the trap for a potential depth of kT for different molecular weight PEO molecules. The dash-dotted line gives the surface area of the event horizon for PEO molecules with molecular weight of $900\,$kDa, the dotted line for $300\,$kDa, and the solid line is for $100\,$kDa.}  
\label{calcplot}
\end{figure}

The thermal energy of a molecule does not depend on its molecular weight. Consequently, the threshold trapping potential necessary for stable trapping is independent of the molecular weight. However, since the achievable trapping potential for a given trapping laser power does depend on the molecular weight, molecules with different molecular weight (and therefore different polarizabilities, see eqn. (2)) have different threshold trapping powers, as confirmed by both our experiments (fig. (3b)) and our calculations (fig. 4). Both our results and calculations also show that for the trapping powers we used ($P<0.7\,$W) the achievable trapping potential for molecules with molecular weight of $m_w=100\,$kDa was not sufficient to stably confine these molecules. 

While it is tempting to conclude that the qualitative agreement between
theoretical expectations and our observations allow us to definitively
identify optical trapping as the mechanism responsible for the increase
in concentration of PEO molecules, it is important to rule out other
possible mechanisms that might contribute. For example, the combination
of convection and thermophoresis was used to Braun and
Libchaber~\cite{braun2002} to trap DNA. It should be noted that the
geometry used by Braun and Libchaber differs significantly from ours ---
their trapping occurred in the bottom 5\,$\mu$m of their sample chamber,
while in our experiments, the molecules are trapped in a volume of
approximately 1\,$\mu$m $\times$ 1\,$\mu$m $\times$ 3\,$\mu$m located
60\,$\mu$m above the bottom of our sample chamber (which had a total
depth of 500\,$\mu$m). Therefore, the interaction between convection,
thermophoresis, and the chamber floor that allowed trapping by Braun and
Libchaber cannot occur in our experiment. As both the objective and
sample slide in our experiment were maintained at a constant
temperature, the only heating would be in the vicinity of the focus,
which is where we observed the increase in concentration. Noting that
the Soret coefficient of PEO in water is positive~\cite{kita2004},
thermophoresis would oppose trapping, and can be discounted as a
mechanism responsible for trapping --- in the absence of the three-way
interaction between temperature gradient, convection, and chamber
bottom, thermophoresis alone would act to \emph{reduce} the
concentration.

We can estimate an upper limit to convective flow in the trap. At the
maximum power we had available, the expected temperature rise, not
accounting for the effect of cooling the sample slide or objective,
would be 13\,K~\cite{peterman2003}. As this heating is localised in the
focal region, we can approximate it as a uniformly heated sphere of
water, surrounded by cooler water. Over-estimating (since we shall be
content with an upper limit) the size of such as sphere as 10\,$\mu$m in
diameter, the net upward force due to the reduced density is 0.019\,pN,
which will be in equilibrium with viscous drag at a speed of
0.2\,$\mu$m/s. The actual flow speed is likely to be much slower (for
example, assuming a diameter of 5\,$\mu$m, also still an overestimate,
gives a speed of 0.06\,$\mu$m/s). In any case, convective flow would be
fastest in the focal volume of the trap, and act against trapping by
pulling particles out of the trap through viscous drag. It is possible
that convective flow might deliver PEO molecules to the trap, but unless
the optical gradient force can exceed the drag, the molecules will not
be trapped.

As both thermophoresis and convection can be ruled out as causes of the
trapping we observed, this leaves optical gradient forces as by far
the most likely mechanism at work. Thus we can, with confidence, state
that we have optically trapped the PEO molecules.

We have shown that using optical tweezers it is possible to locally increase the molecule concentration of large molecules. We propose that this method could be used to assist nucleation of protein crystals. In contrast to conventional methods---where nucleation is triggered by changing the properties of the whole growing medium---the suggested tweezers-assisted nucleation would favor the creation of only one nucleation site in the entire solution. 
The ability to affect molecules increases with molecular weight. Crystals of large proteins are especially difficult to grow, and their structure cannot be analyzed using alternative methods like NMR~\cite{tugarinov2004}. For smaller protein molecules, alternatively, the concentration of a large molecular weight precipitant could be increased, which results in a lower saturation concentration of the respective proteins. 

We would like to acknowledge the support from the University of Queensland and the Australian Research Council.

\end{document}